\documentstyle[twocolumn,prb,aps,psfig]{revtex}
\begin{document}
\draft
\title{Spin-ordering in S = 1 anisotropic Heisenberg models
with nondiagonal spin exchange}
\author{M. Pleimling}
\address{Institut f\"ur Theoretische Physik B, Technische Hochschule,
D--52056 Aachen, Germany}
 
\maketitle
 
\begin{abstract}
The properties of S = 1 anisotropic Heisenberg models with
nondiagonal exchange between axial and planar spin components
are investigated using Monte Carlo techniques. The quantum nature
is taken into account in a semi-classical approximation. The
ordering of the spins when applying an external field with axial
and planar components is discussed. It is argued that the quantum
nature of the spins and the nondiagonal exchange may explain the
peculiar shape of the magnetic specific heat of FeBr$_2$ 
as well as
the weakly first--order phase transition
observed in the same compound when a tilted field is applied.\\
\end{abstract}


\section{Introduction}
Metamagnets of Ising type display a field--induced phase transition
from an antiferromagnetic phase to a (saturated) paramagnetic phase,
the transition being of first order at low temperatures and of second
order at high temperatures.\cite{kin75,str77} The first and second order
transition lines meet at a tricritical point. The layered antiferromagnet
FeCl$_2$ is generally considered to be the textbook example of this
kind of antiferromagnets with a strong uniaxial anisotropy.\cite{bir72}

Recently, there has been a renewed interest in the layered hexagonal
antiferromagnet FeBr$_2$.\cite{dea95,sel95,sel96,aru96,kat97,ple97,hel97,pet97,pet98}
In the ordered
phase of FeBr$_2$ the spins of the iron ions are aligned 
ferromagnetically in the triangular layers, the layers being stacked
antiferromagnetically along the $c$--axis. Adjacent iron layers are
separated by two nonmagnetic bromide planes.

Investigations of de Azevedo et al. \cite{dea95} revealed noncritical
anomalies in the antiferromagnetic phase of FeBr$_2$ below the
transition line to the paramagnetic phase, e.g.\ maxima or shoulders
in the temperature derivatives of the magnetization at constant axial
fields. The ingredients crucial for the existence of these noncritical spin
fluctuations
are the effectively weak ferromagnetic intralayer couplings
and the large number of interlayer nearest neighbors (due to the
superexchange mediated by the bromide planes).\cite{sel95,sel96,ple97}

Surprisingly, measurements of the magnetic specific heat \cite{aru96}
revealed a sharp peak superposed on a broad shoulder or maximum below
the transition to the paramagnetic phase, in contrast to Monte Carlo
simulations of Ising metamagnets \cite{sel95,sel96,aru96,ple97}
showing only a noncritical shoulder or maximum. The peculiar shape
of the specific heat of FeBr$_2$ was tentatively interpreted \cite{aru96,kat97}
as the signature of a new phase transition between two different
antiferromagnetic orderings of the $z$--components of the 
spins. This new transition line was supposed
to result from the decomposition of the tricritical point into a
bicritical point and a critical end point, a possible scenario
emerging from the mean field treatment of Ising metamagnets.\cite{kin75}
Remarkably, however, such a decomposition of the tricritical point
has not been observed in Monte Carlo simulations of three-dimensional
Ising antiferromagnets.\cite{her93,sel96,ren96} Furthermore, there are no
indications of an antiferromagnetic--antiferromagnetic phase transition
in the magnetometric measurements in an axial magnetic field.\cite{dea95,pet97}

Recent investigations in an external field
with an axial and a planar component have shown evidence for 
jumps in the magnetization parallel and perpendicular
to the field.\cite{pet98} 
The authors suggest that this 
transition--like phenomenon may involve a nondiagonal exchange
between the axial and planar components of the spins.

Hitherto, theoretical progress in understanding the properties of
FeBr$_2$ has mainly been achieved by analyzing simplified spin
$\frac{1}{2}$ Ising antiferromagnets.\cite{her93a,sel95,sel96,aru96,ple97}
However, the lowest state of an iron ion in FeBr$_2$ is a triplet consisting
of a lowest doublet and a singlet.\cite{bal85} Therefore, in Ref.\
9, anomalies of the magnetization and the specific heat 
have also been studied in S = 1 models.

In the present work, I shall present a Monte Carlo study of
S = 1 anisotropic Heisenberg models which
include a nondiagonal exchange term between axial and planar spin
components. The quantum nature of the S = 1 spins will be taken into
account in a semi--classical approximation. 
Especially, I shall 
give a possible explanation of the peculiar 
shape of the specific heat of FeBr$_2$ and discuss the 
spin--ordering in an axial and in a nonaxial field.

The paper is organized in the following way: In Section II the 
anisotropic S = 1 Heisenberg models are introduced and 
the semi--classical approximation used in the simulations is 
discussed. Section III deals with the properties of 
ferromagnetic and antiferromagnetic models on a tetragonal lattice.
These simplified models are studied in order to investigate the effects
of (1) the semi--classical approximation, (2) different couplings between
planar and axial spin components, and (3) an external planar field component.
In Section IV a realistic model for FeBr$_2$, both in an axial and in a tilted
field, is discussed and 
the findings are compared to the experiments.
Some of the results of this Section have already been published in a brief
report.\cite{ple98}
A short summary
concludes the paper.

\section{The models}
The low-temperature properties of FeBr$_2$ 
(space group D$_{3d}^3$) may be described by the effective 
S = 1 Heisenberg 
Hamiltonian\cite{yel75,pou93,muk81,ple97}
\begin{eqnarray}
{\cal H} &=& - \sum\limits_{\alpha > \beta }  J_{\alpha \beta } \left\{  \frac{1}{\eta} S_{\alpha}^z 
S_{\beta}^z
+  S_{\alpha}^x S_{\beta}^x + S_{\alpha}^y S_{\beta}^y \right\} \nonumber \\
&& - \sum\limits_{\alpha} D \left( S_{\alpha}^z \right)^2 
- \sum\limits_{\alpha} \left[ 
H_x S_{\alpha}^x + H_z S_{\alpha}^z
\right] + {\cal H}_{nd}(J_{xz})
\end{eqnarray}
defined on a three-dimensional hexagonal lattice consisting of a layered system
of triangular lattice planes stacked along the $c$--axis.
Two different sets of interactions $J_{\alpha \beta}$ have been derived
from inelastic neutron scattering studies:\cite{yel75,pou93} the ferromagnetic
nearest neighbor interaction in the triangular layers perpendicular to the $c$--axis
(corresponding to the $z$--direction), $J_1$, is weakened by 
antiferromagnetic in--layer
couplings, extending either up to next-nearest neighbors,\cite{yel75} $J_2$, or up to
third neighbors,\cite{pou93} $J_3$. Every spin is coupled antiferromagnetically  
with the strength $J'$
to ten spins in the neighboring layer.\cite{her93a} The constant $\eta$ describes
the anisotropy in the exchange interactions. 

The second term in Equation (1) describes a single--ion anisotropy with the easy axis
of the spin along the $z$--axis, i.e.\ $D > 0$, 
whereas $H_x$ and $H_z$ are the planar and axial components of an
applied external field. Finally, the last term, ${\cal H}_{nd}(J_{xz})$,
is a nondiagonal intralayer exchange
between axial and planar spin components with the strength $J_{xz}$,\cite{muk81}
involving only products of pairs of spins.
In order to illustrate this interaction, which is invariant when applying
the symmetry elements of the trigonal point group,
consider two
neighboring in-layer sites, called 0 and 1, with the same value of $y$.
The nondiagonal exchange may be written in the form\cite{muk81}
\begin{equation}
{\cal H}_{nd}(J_{xz}) = \sum\limits_{\left< \alpha \beta \right>} H_{\alpha \beta}(J_{xz})
\end{equation}
where the sum is over nearest-neighbor in-layer sites $\left< \alpha \beta \right>$. For the pair
$\left< \alpha \beta \right> = \left< 01 \right> $ this exchange takes the form\cite{muk81}
\begin{equation}
H_{01} = - J_{xz} \left( S_0^z S_1^x + S_0^x S_1^z \right) .
\end{equation}
The other terms in Equation (2) are obtained by applying the appropriate symmetry
elements transforming the pair $\left< 01 \right>$ onto $\left< \alpha \beta \right>$.

As this model, defined on a hexagonal lattice, is rather complicated, it is preferable
to study the effects of the semi-classical approximation, of the nondiagonal exchange,
and of the planar field component first in simpler models. Therefore, I consider also
the following S $= 1$ Heisenberg Hamiltonian defined on a tetragonal lattice:
\begin{eqnarray}
{\cal H} &=& - J \sum\limits_{ijk} \vec{S}_{ijk} \left( \vec{S}_{i+1 \, j k} + \vec{S}_{i \, j+1 \, k}
\right) - J' \sum\limits_{ijk} \vec{S}_{ijk}  \vec{S}_{ij \, k+1} \nonumber \\
&& - 
D \sum\limits_{ijk} \left( S^z_{ijk} \right)^2 - \sum\limits_{ijk} \vec{H} \vec{S}_{ijk} + 
{\cal H}_{nd}(J_{xz})
\end{eqnarray}
with $\vec{S} = \left( S^x,S^y,S^z \right)$ and $\vec{H} = \left( H_x, 0, H_z \right)$. 
Here $i$, $j$, $k$ (corresponding to the 
a, b, c--directions respectively)
label the lattice sites. The coupling in the layers, $J$, is ferromagnetic, whereas the coupling
between adjacent layers, $J'$, may be antiferromagnetic ($J' < 0$) or ferromagnetic
($J' > 0$). The single-ion anisotropy, $D > 0$, is considered in order to obtain
ground states with nonvanishing $z$--components of the spins.
This may alternatively be achieved by considering instead an anisotropy in the exchange
interactions (yielding three-dimensional XXZ models). 

For the tetragonal models, I will consider two different couplings between planar and
axial spin components: (1) a term involving only products of pairs of spins, being 
invariant under the symmetry operations of the point group $C_{4v}$ (this term can
thus be considered to be analogous to the nondiagonal exchange term, see Equations
(1)-(3), proposed for the description of FeBr$_2$),
\begin{eqnarray}
{\cal H}^1_{nd}(J^1_{xz}) &=& - J^1_{xz} \sum\limits_{ijk} \left[ S^z_{ijk} S^x_{i+1 \, jk}
- S^x_{ijk} S^z_{i+1 \, jk} \right. 
\nonumber \\
&& \left. + S^z_{ijk} S^y_{i \, j+1 \, k} - S^y_{ijk} S^z_{i \, j+1 \, k}
\right],
\end{eqnarray}
or (2) a term involving products of squares of spins, being invariant under the symmetry operations
of the space group,
\begin{equation}
{\cal H}^2_{nd}(J^2_{xz}) = - J^2_{xz} \sum\limits_{\left< \alpha \beta \right>} 
\left( S^z_{\alpha} \right)^2 \left[ \left( S^x_{\beta} \right)^2 + \left( S^y_{\beta} \right)^2
\right],
\end{equation}
where the sum is over nearest--neighbor in--layer sites.

I consider in the following S = 1 spins where the quantum nature
is treated in a semi--classical approximation: the $z$--component is discretized
and can only take the values 1, 0, or $-1$, whereas the spin length is fixed to be
$\left| \vec{S} \right| = \sqrt{\mbox{S} \left( \mbox{S} + 1 \right)} = \sqrt{2}$.
Hence, the spins rotate in the $xy$ plane like a classical vector. The $xy$--components
provide additional energy contributions as compared to the S = 1 Ising model,
even at low temperatures. Note that in Ref.\ 9 the spin length was fixed at 1,
the spins thus having a planar component only if $S^z = 0$, yielding
thermal properties similar to the S = 1 Ising model.

The ground states of the semi--classical antiferromagnetic 
hexagonal and tetragonal models
are readily obtained provided the nondiagonal exchange 
constant is not too large.
In small axial fields ($H_x = 0$ and $H_z < H_z^c = -2 N J'$, $N = 10$ for 
the hexagonal model and
$N = 2$ for the tetragonal model), the antiferromagnetic phase,
where the spins are aligned ferromagnetically in the layers, with
an antiferromagnetic arrangement between subsequent layers, is stable. For larger
fields ($H_z > H_z^c$ and $H_x=0$) the axial spin components $S^z$ order
ferromagnetically, whereas the signs of the
planar components ($S^x$ and $S^y$) still change from layer to layer.
Note that in absence of an ordering planar field the ground state
is infinitely degenerate, as the angle between the planar spin components and,
say, the $x$--axis is not fixed.
Keeping $H_z$ constant and applying a planar field, $H_x$, yields a stable phase
where the $xy$--components of the spins are ordered in a spin--flop phase, in which
the magnetization per layer in $x$--direction has the same value in each layer,
whereas the $y$--components have different signs in adjacent layers but the same
absolute value. For stronger planar fields the $y$--components finally vanish,
leading to a ferromagnetic ordering of the planar spin components.

The effect, at $T=0$, of a nondiagonal coupling between spin components
can be discussed for the tetragonal ferromagnetic model. For small absolute values 
of the nondiagonal exchange constant (be it $J^1_{xz}$ or $J^2_{xz}$,
see Equations (5) and (6)), the ground 
state is given by a ferromagnetic phase where the spins have both axial and planar
components. Consider first the two--spin exchange with the coupling constant $J^1_{xz}$. 
Increasing $\left| J^1_{xz} \right|$ does not change the energy of the ferromagnetic
phase, whereas the energy of a second phase, consisting of ferromagnetically coupled 
layers with a stripped pattern in every layer, is decreased, until, for
$\left| J^1_{xz} \right| > \sqrt{2} J +\frac{\sqrt{2}}{4} H_x$, the stripped phase has,
at $T=0$, a lower energy than the ferromagnetic phase, thus forming the ground state.
Note that the appearance of a new ground state at the threshold value does
not depend on the sign of $J^1_{xz}$.
If the second spin exchange with negative coupling constants $J^2_{xz}$ is considered,
the energy of the ferromagnetic phase is increased when $J^2_{xz}$
decreases. For couplings $J^2_{xz} < - \frac{D + H_z + \left( 1 -\sqrt{2} \right) H_x}{4}$
the ground state is given by a ferromagnetic phase where the spins have only a
planar component. A positive coupling constant $J^2_{xz}$
does not yield a new stable phase. The ground state of the tetragonal 
antiferromagnetic model in presence of a nondiagonal spin exchange can be
discussed in similar terms. In the following it is always supposed that the 
nondiagonal coupling constant does not exceed the threshold value which leads
to a new ground state.

The thermal properties of the different models
were investigated by simulating systems consisting
of $L \times L \times L$ Heisenberg spins, with $L$
ranging from 10 to 30.
As the equilibration proved to be rather slow, the first
$3 \times 10^4$ Monte Carlo steps per site were usually discarded. Simulations for
up to 20 different realizations with different random numbers were performed 
in order to improve
the statistics. Besides the energy $E$ and the specific heat $C$, the different
components of the magnetization per layer, $m_x(i)$, $m_y(i)$, and $m_z(i)$, and,
for the antiferromagnetic models,
related quantities such as the sublattice magnetizations, $\vec{M}_1 =
\left( m_x^1, m_y^1, m_z^1 \right)$ and $\vec{M}_2 =
\left( m_x^2, m_y^2, m_z^2 \right)$, referring to odd and even layers, were 
computed. Here, the odd, respectively even layers have the magnetization
$m_z^1 = +1$, respectively $m_z^1 = -1$ at $T=0$. The applied axial field points
in the "+"--direction, i.e.\ it tends to destabilize the even layers.

The ground state was always used as starting configuration for the simulations.
In absence of a planar field component, the state with $S^x = 0$ was
chosen among the infinity of degenerate ground states.
Note that in this case one encounters, after initial relaxation, a metastable
state, in which the system remains, possibly, for a long time, the time depending
on the size of the system and the temperature. The computed planar spin 
components are then supposed to be very close to their values in the 
thermodynamic limit.

\section{Thermal properties of the tetragonal models}
In this section, the thermal properties of the tetragonal ferromagnetic
and antiferromagnetic models (see Equation (4)) are discussed. For a 
vanishing external field, the Hamiltonian of the antiferromagnetic
model can be mapped in the usual way onto that of the ferromagnetic model.
Therefore, I will in the following present for $H_x = 0$
simulations of the ferromagnetic model, the antiferromagnetic model
being analyzed for $H_x \neq 0$.

Figure 1 shows the temperature dependent specific heat and the
magnetization obtained for vanishing nondiagonal spin exchange 
for the ferromagnetic model, 
with $J = J'$, $D = 3 J'$,
and $H_x = 0$. The
specific heat has a two peak structure, see Figure 1a: the peak at $T_c$
is the critical peak resulting from the disordering of the $z$--components
of the spins, whereas the low--temperature peak at $T_{xy}$ follows from the
disordering of the planar spin components. The magnetization data, see 
Figure 1b, show that the $z$--components, in absence of a coupling between planar
and axial spin components, are not affected by this disordering.

This two  peak structure of the specific heat should be compared to
the specific heat of the classical anisotropic Heisenberg model showing
only one maximum. For vanishing single--ion anisotropy the two peaks 
of the specific heat of the semi--classical model merge
to a single peak located at the critical temperature of the
corresponding classical isotropic Heisenberg model.

It is clear from Figure 1 that, due to the discretization of $S^z$
and the presence of the single--ion anisotropy, the disorderings
of the axial and planar spin components are largely decoupled.
The $z$--components behave like
S = 1 Ising spins and disorder at the critical temperature $T_c$ (which increases
with increasing value of $D$, yielding in the limit
$D \longrightarrow + \infty$ the critical temperature of the 3d Ising model), whereas the
xy--components form a classical three-dimensional plane-rotator disordering at the temperature\cite{Fer73}
$T_{xy} = 2.2 J'$ (setting the Boltzmann constant equal to one). 

Note that the specific heat does not vanish when $T$ approaches 0.
This is an artefact of the classical nature of the planar spin components
yielding $\lim_{T \to 0} C = \frac{1}{2}$.

In presence of the nondiagonal two--spin exchange with coupling
constant $J^1_{xz}$ (Equation (5)), the ordering temperature $T_c$
of the axial spin components is decreased, see Figure 2. This results from the coupling,
at temperatures above $T_{xy}$, of the ordered $z$--components to the disordered
planar components, yielding additional fluctuations which lead
to a decrease of the ordering temperature of
the axial spin components. The value of $T_{xy}$
is not affected by this nondiagonal spin--exchange, as long as 
$\left| J^1_{xz} \right|$ is not too large, see below. The spin exchange involving
products of squares of spins (Equation (6)) 
also leads to a decrease of $T_c$ for negative values
of $J^2_{xz}$. This second coupling is not so effective in destabilizing the
ferromagnetically ordered $z$--components, yielding a smaller decrease of $T_c$.

If the strength of $\left| J^1_{xz} \right|$ is increased beyond a threshold value
(being around 0.65 $J'$ for $J = J'$ and $D= 3 J'$), a jump in the axial component of the
magnetization, $M_z$, is observed, see Figure 3a. This discontinuous change, which
is induced by the disordering of the planar components, shows up as a sharp peak in
the specific heat (Figure 3b). At temperatures slightly above $T_{xy}$ some 
orderness of the $z$--components still persists, i.e. $M_z \neq 0$. Increasing further the
temperature, $M_z$ increases until, at $T_c$ (= 3.1 $J'$ for the parameters
of Figure 3), the disordering of the $z$--components finally takes place, 
yielding a second peak in the specific
heat. This behaviour of $M_z$ may be better understood when the temperature, 
starting from the disordered high temperature phase, is decreased.
At $T_c$, the $z$--components order at the usual second order phase
transition. 
The increase of the ordering of the $S^z$ spins at temperatures below $T_c$
increases the effect of the nondiagonal exchange, yielding, after an
initial slow rising, effectively a decrease
of $M_z$, until, at $T_{xy}$, the planar components order. Note that in this case
both $T_c$ and $T_{xy}$ are shifted to lower temperatures, the shift being larger
for larger values of $\left| J^1_{xz} \right|$.

Interestingly, no jump in $M_z$ occurs when the second spin exchange with
coupling constant $J^2_{xz}$ is considered. For all values of $J^2_{xz}$
yielding the ferromagnetic ground state with both planar and axial spin
components, see above, $M_z$ changes continuously at $T_{xy}$.

Figure 4 shows the specific heat of the antiferromagnetic tetragonal model  
obtained when planar fields with different
strengths are applied and only diagonal couplings are considered. 
Increasing the value of $H_x$ moves $T_{xy}$ to lower
temperatures, whereas $T_c$ is not changed. In presence 
of a planar field unusual strong
finite--size effects are observed when the planar spin components
disorder, making it necessary to simulate at least systems of length
$L = 20$ in order to get a reliable estimate of $T_{xy}$. 
These strong finite--size effects are due to the presence of the
$S^z = 0$ states and do not show up for the corresponding
classical plane--rotator,
as I checked. 
Finally, one should notice that, in presence of the nondiagonal
coupling (5) with the strength $J^1_{xz}$, the total magnetization in $z$--direction 
again changes discontinuously for values of
$\left| J^1_{xz} \right|$ larger than the threshold value, which
seems
not to depend on the value of $H_x$.

\section{Thermal properties of the hexagonal model}
Most of the features discussed in the previous Section for the tetragonal
models are also encountered if the realistic hexagonal model (1)
for FeBr$_2$ is considered.\cite{ple98} In the following I will
not reiterate the discussion on the values of the diagonal couplings of the
spins (the interested reader is referred to Ref.\ 9) and
choose a set of parameters, based on inelastic 
neutron scattering experiments\cite{yel75,pou93},
for which pronounced anomalies in the magnetization data and in
the specific heat data are observed\cite{ple97}: 
$J_1 = -16.75 J'$, $J_2 = 0$, $J_3 = -0.29 J_1$, and $\eta = 0.78$.

Figure 5 shows the specific heat and the sublattice magnetizations obtained
for strong single--ion anisotropy $D$ and vanishing nondiagonal interactions,
with $H_z = -18 J'$ and $H_x = 0$.
The temperature dependent specific heat $C(T)$ 
has a three peak structure: the peak at $T_c$
is the critical peak resulting from the disordering of the $z$--components of the
spins. Going to lower temperatures, one encounters first the anomaly due to the
noncritical spin fluctuations. This peak does not correspond to a sharp phase
transition and does not show a significant size dependency, in contrast to the
critical peak at higher temperatures.\cite{ple97} Both peaks are present when
computing the specific heat of Ising metamagnets and involve solely $S^z$.
The third peak, again, results from the
disordering of the planar spin components, see Figure 5b.

The presence of the nondiagonal spin exchange moves both 
$T_c$ and the anomaly to lower temperatures,
see Figure 6a, whereas $T_{xy}$ only decreases, moderately, for large values
of $J_{xz}$. The anomaly in the specific heat approaches the $xy$--peak,
its height being reduced, when $J_{xz}$ is increased, and finally 
merges with the $T_{xy}$ peak.
Again, a threshold value exists for $\left| J_{xz} \right|$, at about 18 $J'$, beyond
which the total magnetization in $z$--direction, $M_z = \left( m_z^1 + m_z^2 \right) /2$,
changes discontinuously, see Figure 6b. The
disordering of the planar components at $T_{xy}$ is then also clearly discontinuous,
as can be seen in the jump of the order parameter $M_y^{op} = \left( m_y^1
- m_y^2 \right) /2$, see Figure 6c.

Changing the value of the axial field $H_z$ does not change $T_{xy}$ if
$J_{xz} = 0$. For nonvanishing nondiagonal couplings, however,
the position of the $xy$--peak is shifted to lower temperatures when
the axial field strength is increased.

Figure 7 shows the influence of the single-ion anisotropy $D$ on the 
specific heat for $J_{xz}=0$. Decreasing $D$, the $T_c$ peak and the anomaly are
moved to lower temperatures, whereas $T_{xy}$ is slightly increased. 
A similar behavior is found for $J_{xz} \neq 0$, as shown in Figure 8 for
$J_{xz}= 16.2 J'$ and $D = -8.1 J'$.
For this choice of the 
parameters, the noncritical spin fluctuations appear
at lower temperatures than the disordering of the planar spin components.
The resulting specific heat has a peculiar shape
consisting of a broad shoulder, the anomaly, and a superposed
peak, the $T_{xy}$ peak. This shape is reminiscent of the magnetic specific
heat of FeBr$_2$ and will be discussed in more details below.

Applying an additional planar field component, $H_x$,
leads to a spin--flop phase 
in the $xy$--components at $T = 0$:
the $y$--components of the layer magnetization have opposite signs in the
different sublattices but the same absolute value, whereas the $x$--components are
the same in every layer. As the axial field tends to stabilize the odd or "+" layers
and to destabilize the even or "$-$" layers, the absolute values of $m_y^1$ and $m_y^2$
are close but not identical at temperatures $T > 0$.

Figure 9 shows the different sublattice magnetizations for $J_{xz}= 16.2 J'$,
$D = -8.1 J'$, and $H_x =0.75 H_z$, i.e.\ a field with axial and planar 
components is considered. At low temperatures, 
the antiferromagnetic ordering
of the axial spin components and the spin--flop ordering of the planar spin
components is clearly seen. When the system is heated, a drastic change in the y-component 
of the layer magnetization takes
place at $T_{xy}$ well below $T_c$. For the chosen parameters, the change in
the $y$--components is continuous. For values of
$\left| J_{xz} \right|$ larger than the threshold value, e.g.\ $J_{xz}= 21.6 J'$, 
this change occurs through 
a jump in the layer magnetization leading to a first--order transition.
At first it may be surprising that the $y$--components (and also the
$x$--components) of the sublattice 
magnetization are not equal at temperatures above $T_{xy}$ and below $T_c$, see
Figure 9. In fact, this is explained by the finite value of the axial field
$H_z$ which tends to destabilize the $z$--components in the "$-$" layers, thus leading
to different sublattice magnetizations in the different layer types. As a result
the $y$--components do not vanish at the phase transition at $T_{xy}$,
but take a small value which is different for the two layer types. Only
when the $z$--components of the sublattice magnetizations are equal, at $T_c$,
do the $y$--components vanish and the $x$--components take the same value. 
This effect is also observed when applying a field with axial and
planar components to the tetragonal antiferromagnetic model discussed
in the previous Section.

Increasing the value of the planar field
moves $T_{xy}$ to lower temperatures,
as it is also the case for the tetragonal antiferromagnetic 
model. 
Furthermore, the specific heat seems to loose its peculiar shape for large values of $H_x$.
At least for the temperature resolution used in the present study,
only one, rather broad, peak is seen instead of the shoulder or maximum with a
superposed sharp peak which is observed for $H_x = 0$. 

The presented 
data clearly indicate that the quantum nature of the 
S = 1 spins has to be taken into account in a theoretical description of 
the low temperature behaviour of FeBr$_2$.
The semi--classical approach adopted in the present work leads to different
disordering temperatures for the planar and the axial spin components.
Hence, the anomaly and the low-temperature $xy$ peak are not intimately related.
The modification of, for instance, the exchange $J_{xz}$ 
or the degree of Ising-like anisotropy $D$
changes their respective positions and may lead 
in an axial field to specific heat data 
having the peculiar shape
of the magnetic specific heat of FeBr$_2$, see Figure 8. One should notice that,
in order to obtain this peculiar shape with the considered set of
diagonal coupling constants, the single--ion anisotropy has to
be decreased considerably as compared to the values obtained from inelastic neutron
scattering measurements. Nevertheless, one must keep in mind that both $D$
and the strengths of the diagonal couplings
were derived from the experiments without taking a possible
nondiagonal spin exchange into account.\cite{yel75,pou93}

The evolution of the temperature dependent specific heat of FeBr$_2$ with
increasing axial fields is also of interest.\cite{aru96}
For small fields
only one peak, the critical peak at $T_c$, has been observed. For larger fields,
a shoulder with a superposed peak appears at lower temperatures. 
The shoulder evolves into a maximum with increasing fields whereas the
superposed peak becomes sharper, the whole being shifted to lower temperatures. 
This large--field behavior is well rendered
in the simulations, as shown in Figure 10. Indeed, the shoulder, resulting from the
noncritical spin fluctuations, changes into a maximum
when $H_z$ is increased, whereas the $xy$--peak
becomes sharper, the anomaly in the axial spin components
and the disordering of the planar spin components both moving to lower
temperatures. Note, however, that in the present model the disordering
of the $xy$--components also takes place for small
fields, see above.

The magnetization data resulting from the Hamiltonian (1) also compare favorable
to the experiments. For example, without an applied planar field,
the transverse spin--ordering at $T_{xy}$ does not lead to a discontinuity of 
the axial magnetization (if $J_{xy}$ is not too large, see above), in accordance
with the magnetometric measurements\cite{dea95}, showing no jump in the axial
magnetization despite a sharp peak in the specific heat.
An additional  planar field component yields a 
low-temperature spin--flop phase in the $xy$--components, which
is compatible with the experiment.\cite{pet98} The change in the $y$-component in the
magnetization per layer
may then occur through a continuous or a discontinuous phase
transition, depending
on the values of the parameters. Experimentally, one observes, in a tilted field, a 'weakly
first--order' transition,\cite{pet98} thus being supposedly at the border
between these two scenarios.

One should bear in mind when applying the present results to FeBr$_2$ that the 
Hamiltonian (1) may not be complete. Indeed, the recent measurements in
an nonaxial field\cite{pet98} suggest the presence of in--plane anisotropy
in FeBr$_2$, resulting, for example, from magnetoelastic couplings. Thus an
anisotropy is not contained in the present Hamiltonian. Furthermore,
one should recall that the treatment of the Heisenberg Hamiltonian
is only approximative, due to the semi--classical approach adopted
in Section II, leading to a strong decoupling between planar and axial 
spin components.  

Nevertheless, the present study clearly shows the importance of the nondiagonal
spin exchange and of the quantum nature of the S = 1 spins in describing
the low-temperature properties of FeBr$_2$. Especially, the peculiar shape of
the specific heat in an axial field may be 
traced back to a phase transition of the planar spin 
components, taking place, rather by chance, close to the anomaly of the
axial spin components in the antiferromagnetic phase of FeBr$_2$. 

\section{Summary}
Recent experimental investigations showed some intriguing transition--like phenomena
in the magnetization and in the specific heat of the layered antiferromagnet FeBr$_2$.
The observed transverse spin--ordering in a nonaxial field suggests the existence
of a nondiagonal exchange between the planar and the axial spin components.\cite{pet98}

Motivated by these experiments, the properties of S = 1 anisotropic Heisenberg models
(both on a tetragonal and on a hexagonal lattice)
including nondiagonal spin exchange were investigated by Monte Carlo simulations.
The quantum nature of the spins were taken into account in a semi--classical approximation,
where $S^z$ was discretized and the spin length fixed to be 
$\left| \vec{S} \right| = \sqrt{\mbox{S} \left( \mbox{S} + 1 \right)}$. Applied
external fields with axial and planar components were considered. 

These simulations showed that, besides the disordering of the axial spin components
at high temperatures, a second phase transition involving the planar spin
components takes place at lower temperatures. For the tetragonal models, the
effects of two different nondiagonal spin exchange terms were discussed.
It was shown that the nondiagonal coupling involving only products of two spins
could induce, for large values of the coupling constant,
a discontinuous change of the total axial magnetization at
the ordering temperature of the planar spin components, whereas $M_z$ always
changed continuously when the second exchange term involving products of squares
of spins was considered.
 
For the realistic antiferromagnetic model
defined on a hexagonal lattice, which is supposed to describe the
low--temperature properties of FeBr$_2$, the jump in the total
magnetization in axial direction for strong nondiagonal couplings 
was always mirrored by a corresponding jump in the planar sublattice magnetizations.
Furthermore, it was shown that the 
anomaly of the axial spin components
in the antiferromagnetic phase\cite{sel95,sel96,ple97}
and the low--temperature peak resulting from the disordering 
of the planar spin components are not intimately related.
In fact, the data suggest that the transition--like features in the specific heat
(superposed peak on a shoulder or maximum) and in the magnetization (jumps in 
the magnetization observed in a tilted field) of FeBr$_2$ 
are well described if
a nondiagonal coupling in the spin components and the quantum nature of the
spins are taken into account. It seems, in the light of the present
study, that the anomaly in the axial spin components in the antiferromagnetic
phase and the disordering of the planar spin components
occur in FeBr$_2$ closely one to the 
other rather by chance, yielding, for instance, the measured peculiar
shape of the magnetic specific heat.

\acknowledgments
It is a pleasure to thank W.\ Selke for useful and stimulating discussions.

\begin{figure}
\centerline{\psfig{figure=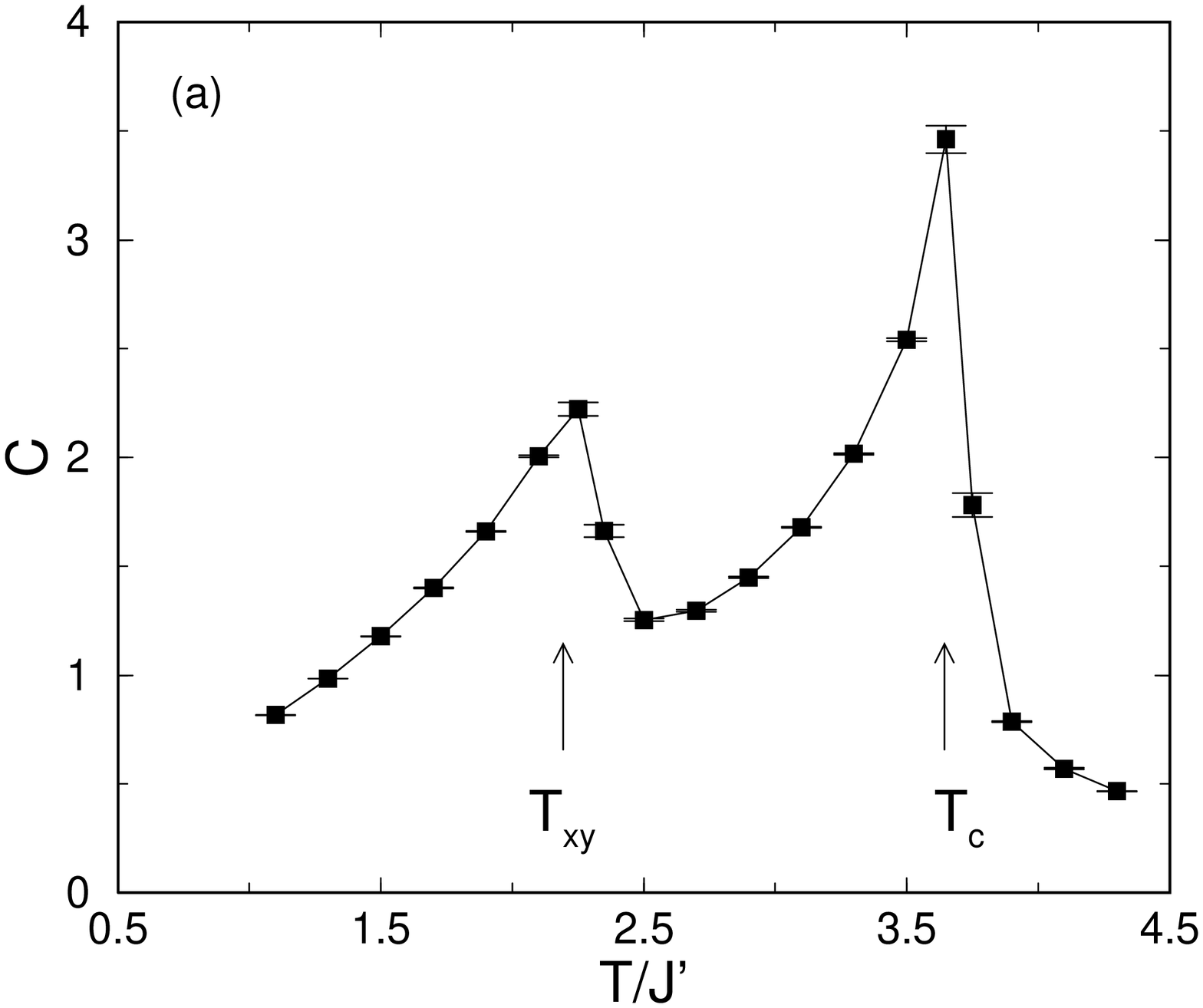,width=7.0cm,angle=270}}
\centerline{\psfig{figure=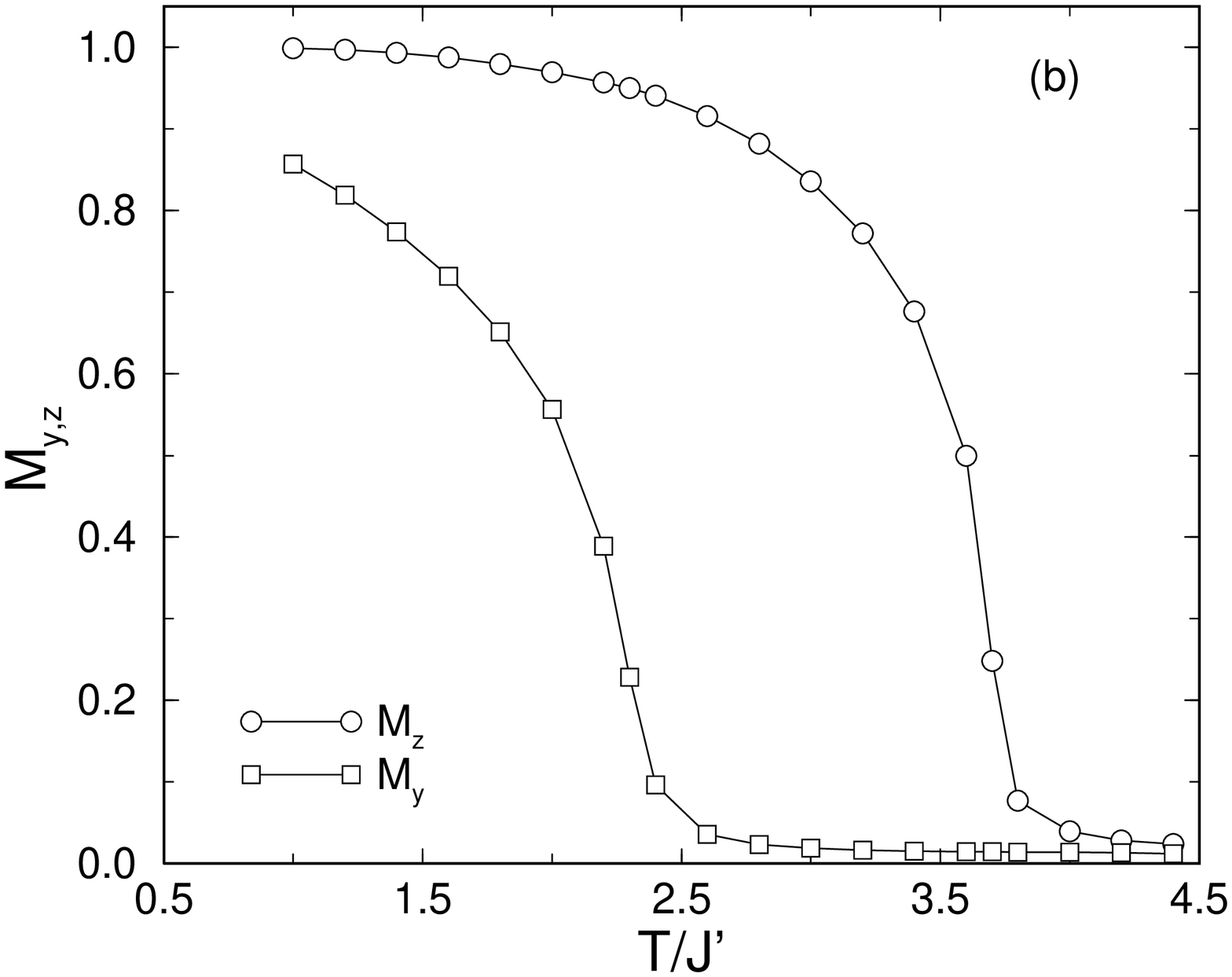,width=7.0cm,angle=270}}
\caption{Monte Carlo data of (a) the specific heat and (b) the components of the
magnetization as function of the temperature
for the S = 1 ferromagnetic Heisenberg model on a tetragonal lattice, with
$J = J'$, $D = 3 J'$, and $H_x = H_z =0$. The system size is $L = 20$.
Here and in the following figures, the Boltzmann constant is set equal to one.
}
\label{fig1} \end{figure}

\begin{figure}
\centerline{\psfig{figure=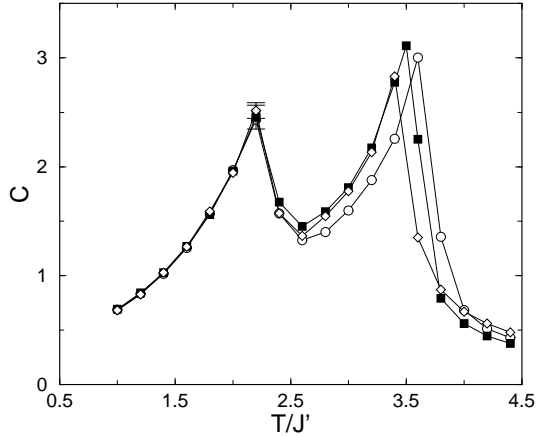,width=7.0cm,angle=270}}
\caption{Specific heat $C$ of the ferromagnetic model
as a function of temperature for the two different
nondiagonal spin exchange terms given in Equations (5) (coupling 
constant $J^1_{xy}$) and (6) (coupling constant $J^2_{xy}$), with 
$J = J'$, $D = 3 J'$, and $H_x = H_z =0$. Open diamonds: $J^1_{xy} = 0.5 J'$.
Filled squares: $J^2_{xy}= 0.5 J'$. The data obtained without a nondiagonal
coupling (open circles) are included for comparison. 
Systems with $10^3$ spins are considered.
Only error bars larger than the size of the symbols are shown.
}
\label{fig2} \end{figure}

\begin{figure}
\centerline{\psfig{figure=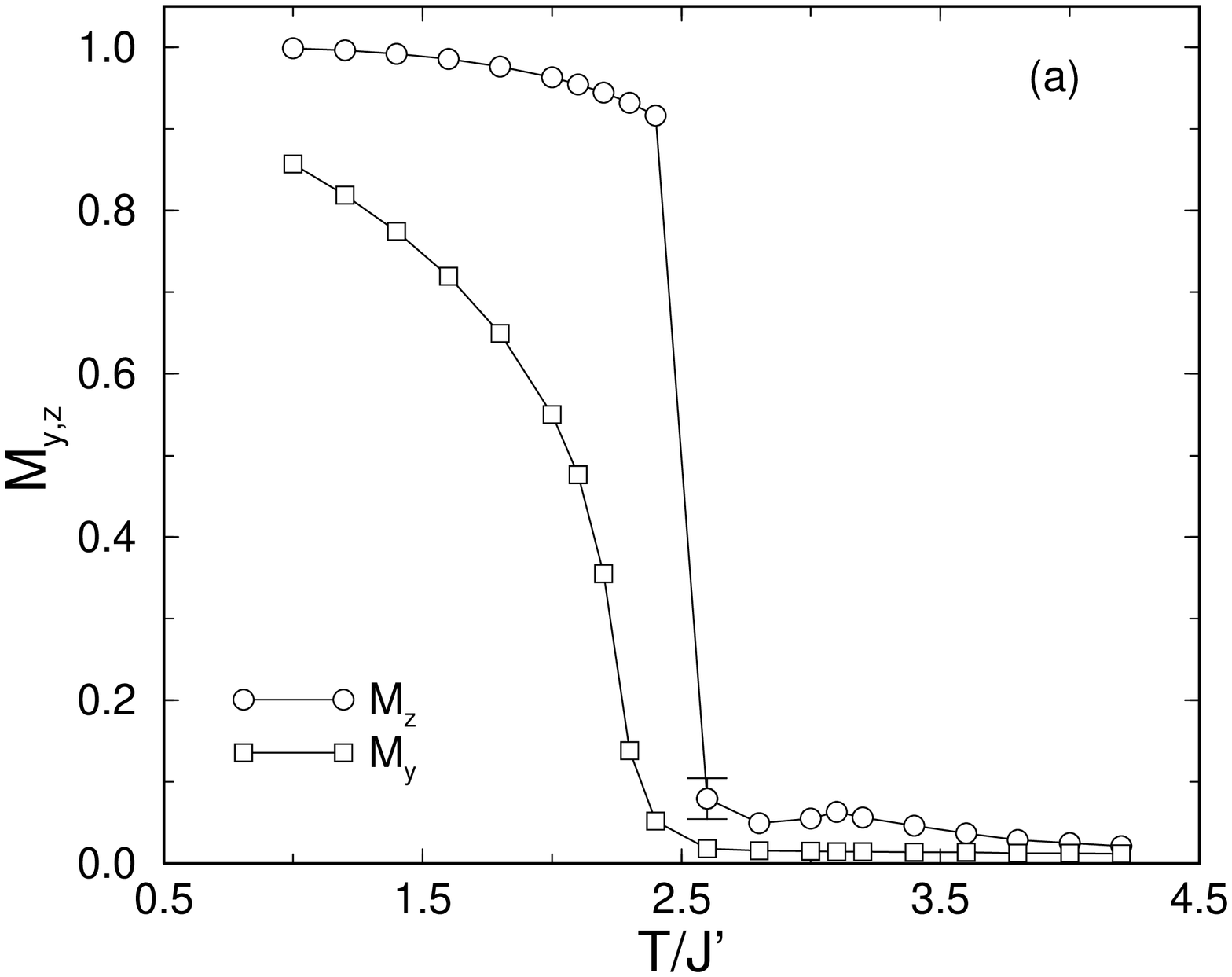,width=7.0cm,angle=270}}
\centerline{\psfig{figure=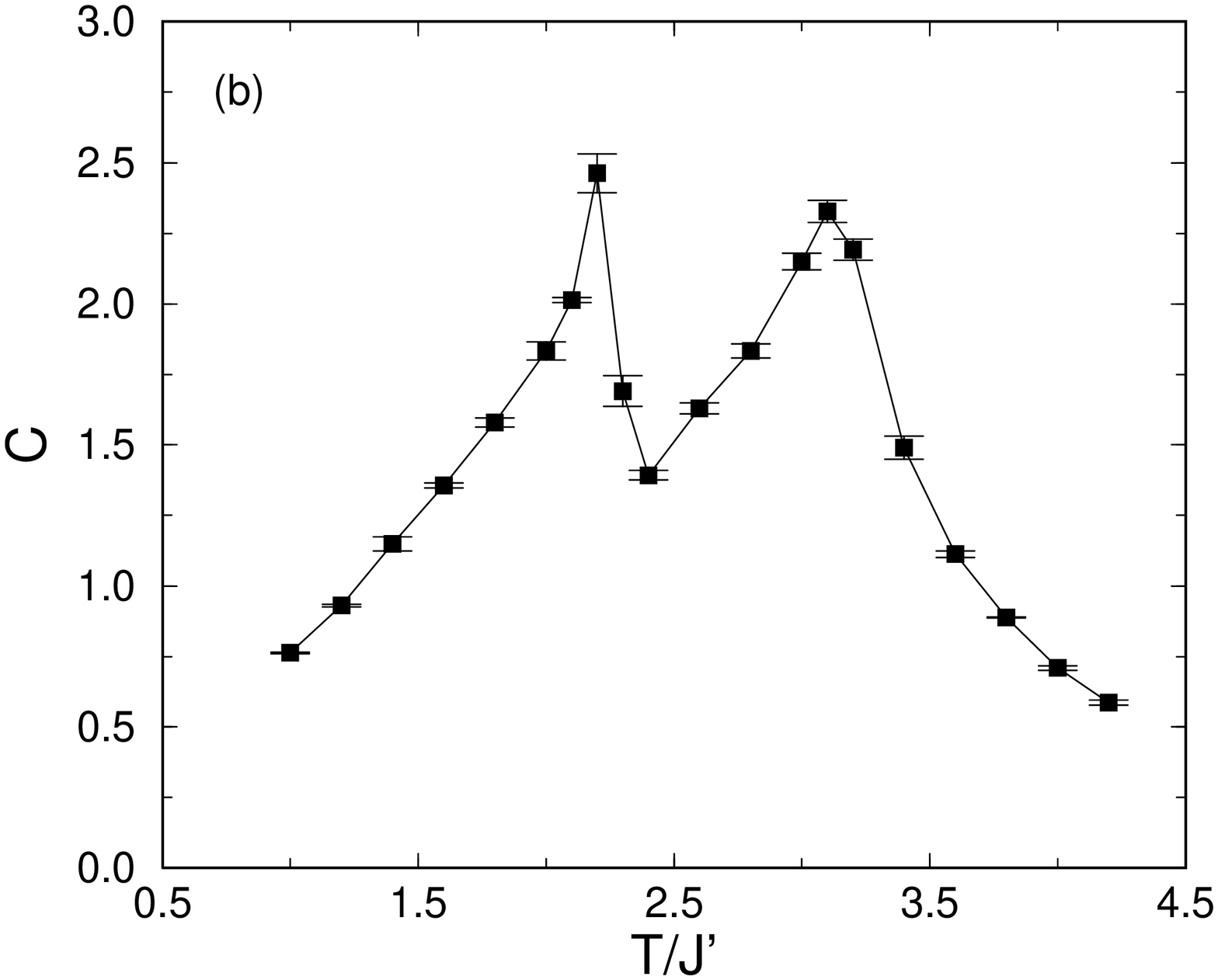,width=7.0cm,angle=270}}
\caption{Monte Carlo data of (a) the components of the magnetization 
and (b) the specific heat as function of the temperature obtained for the
ferromagnetic model on a tetragonal lattice, with 
$J = J'$, $D = 3 J'$, $H_x = H_z =0$, and $J^1_{xz} = 0.7 J'$ . Systems with 
$20 \times 20 \times 20$
spins are simulated.
}
\label{fig3} \end{figure}

\begin{figure}
\centerline{\psfig{figure=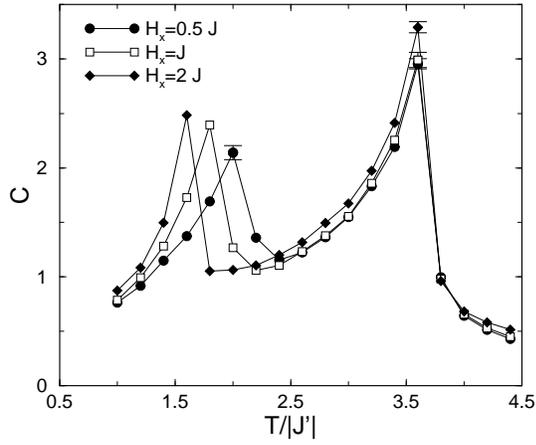,width=7.0cm,angle=270}}
\caption{The temperature dependent specific heat obtained for the 
antiferromagnetic S = 1 Heisenberg model defined on a tetragonal lattice
when planar fields with different strengths are applied,
with $J = - J'$, $D = - 3 J'$, and $H_z =0$. Nondiagonal couplings
between axial and planar spin components are not considered.
The system size is $L = 20$.
Only error bars larger than the size of the symbols are shown.
}
\label{fig4} \end{figure}

\begin{figure}
\centerline{\psfig{figure=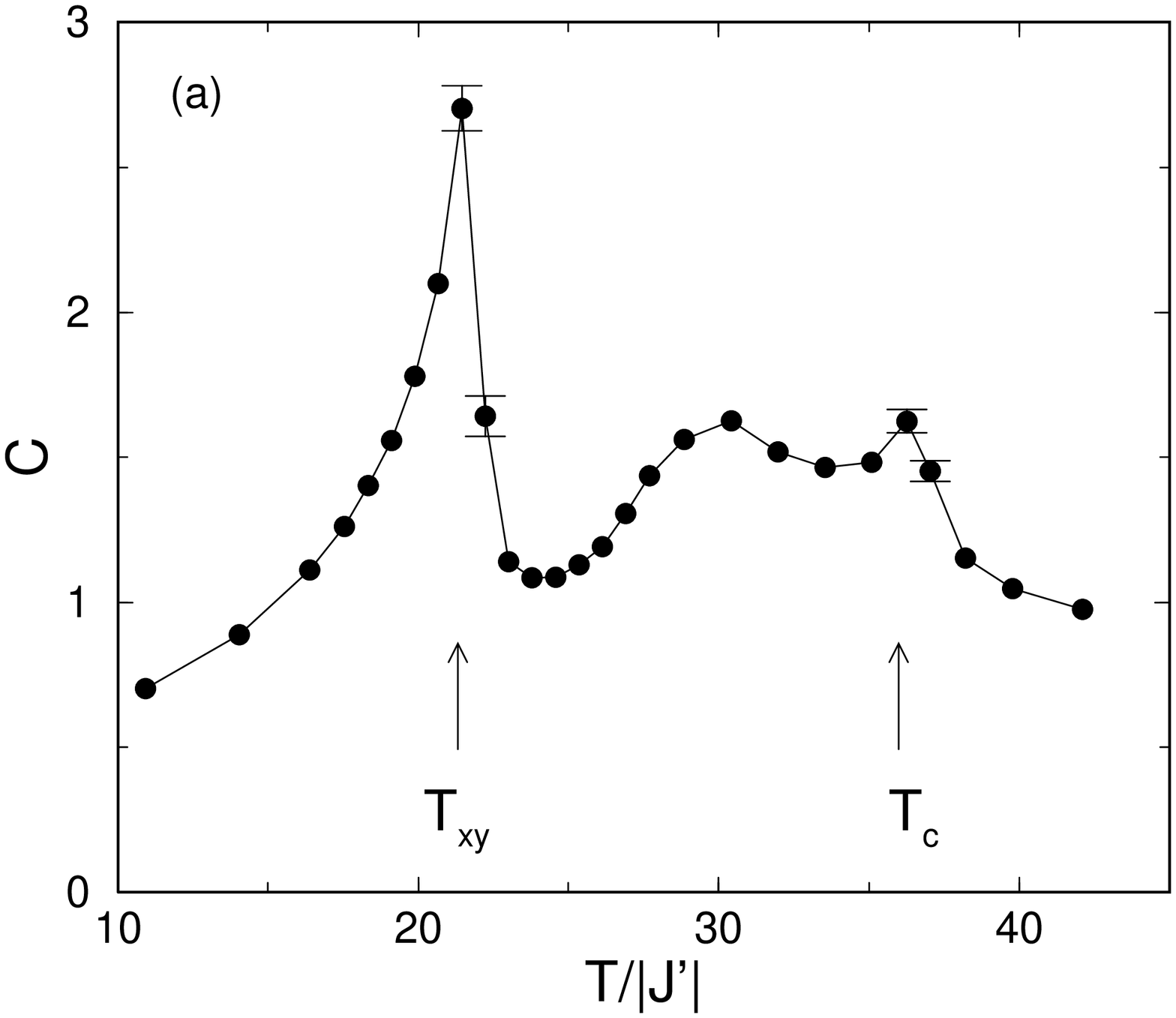,width=7.0cm,angle=270}}
\centerline{\psfig{figure=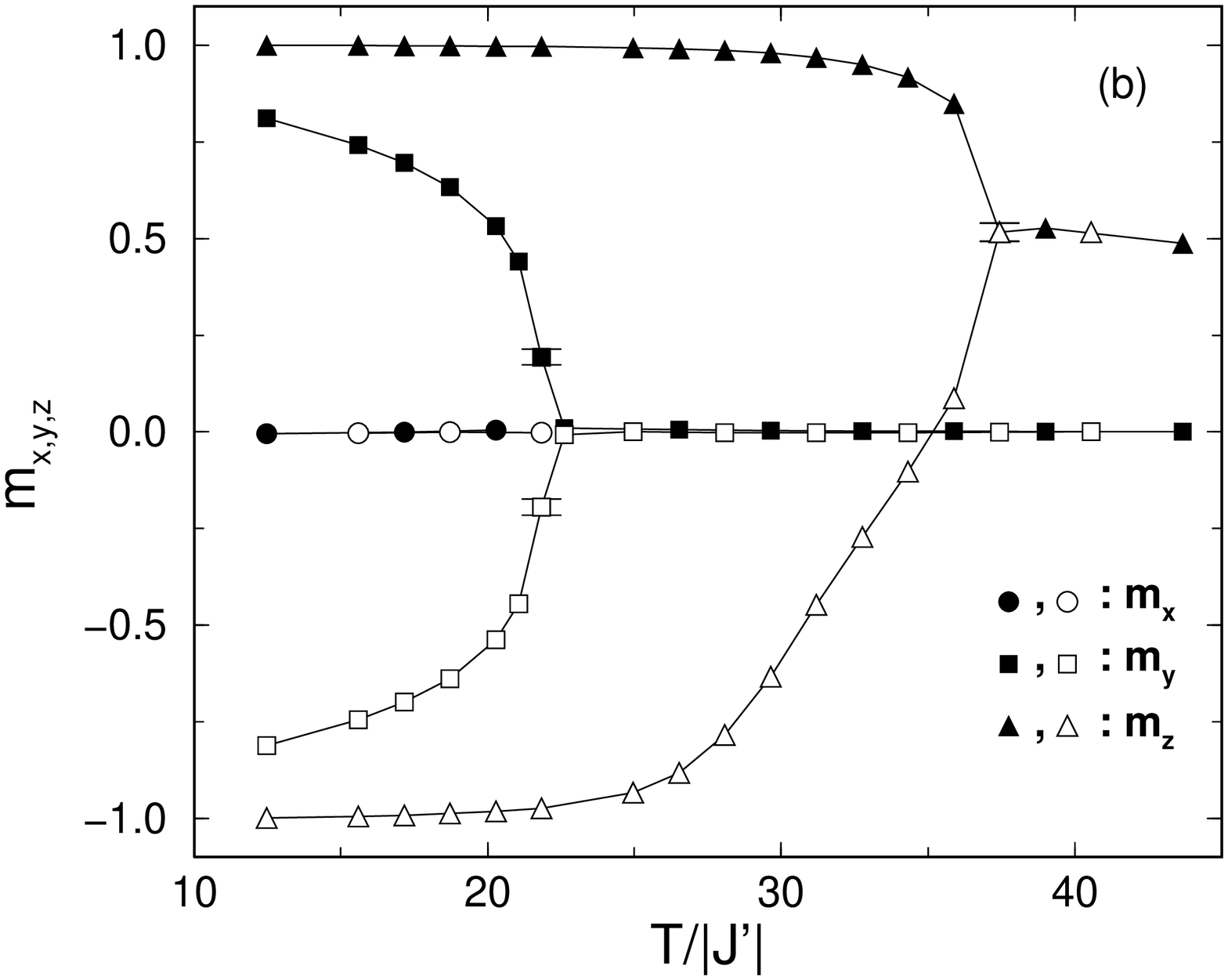,width=7.5cm,angle=270}}
\caption{Monte Carlo data of (a) the specific heat and (b) the components of the 
magnetization per layer in odd (full symbols) and even
(open symbols) planes as function of the temperature
for the S = 1 anisotropic Heisenberg model on the hexagonal
model in an axial field $H_z= -18 J'$, with 
$D = -32.4 J'$ and $J_{xz} = 0$. Systems with $20 \times 20 \times 20$ 
spins are simulated.
Error bars are only shown when they are larger than the sizes of the symbols.
}
\label{fig5} \end{figure}
 
\begin{figure}
\centerline{\psfig{figure=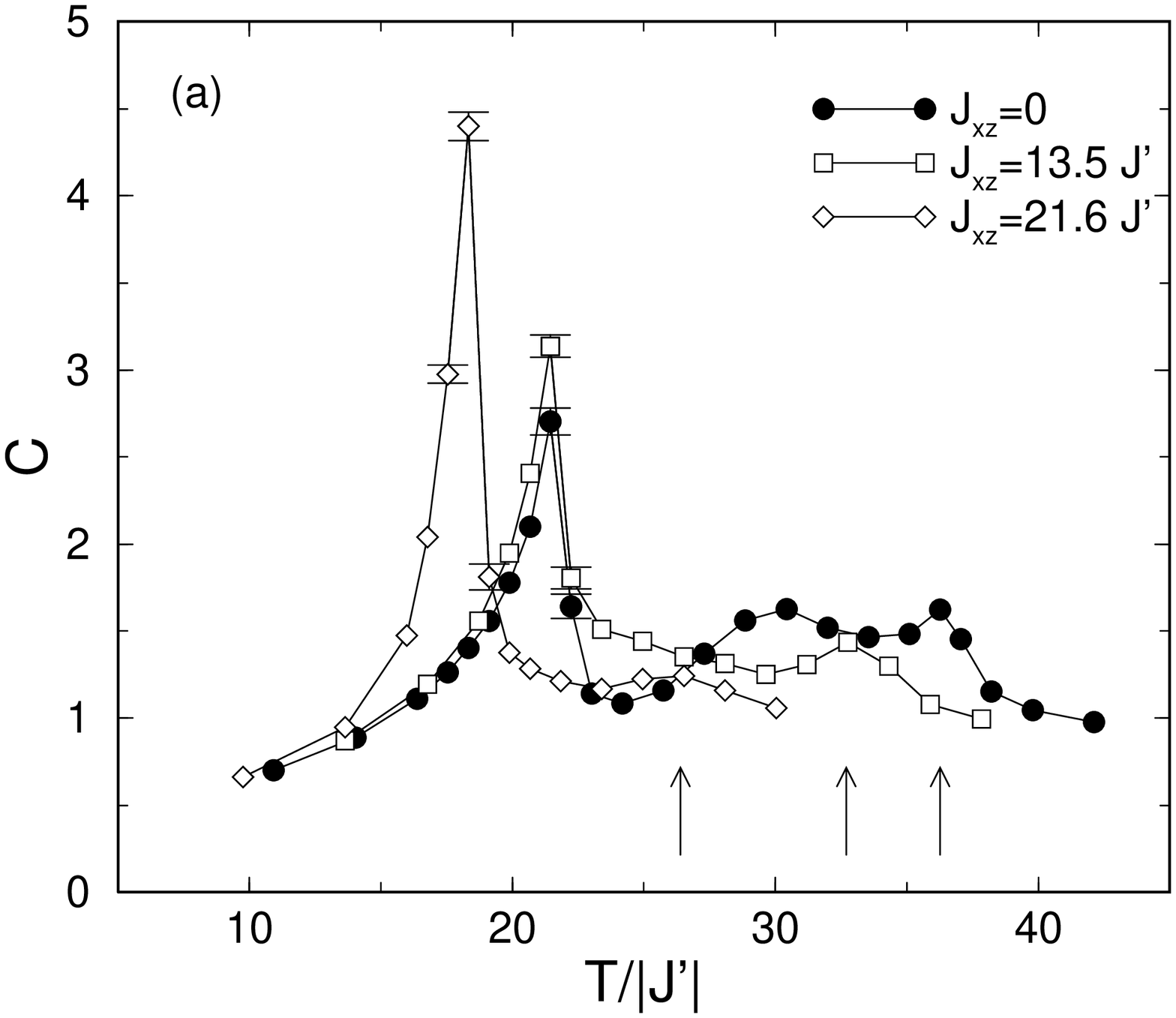,width=7.0cm,angle=270}}
\centerline{\psfig{figure=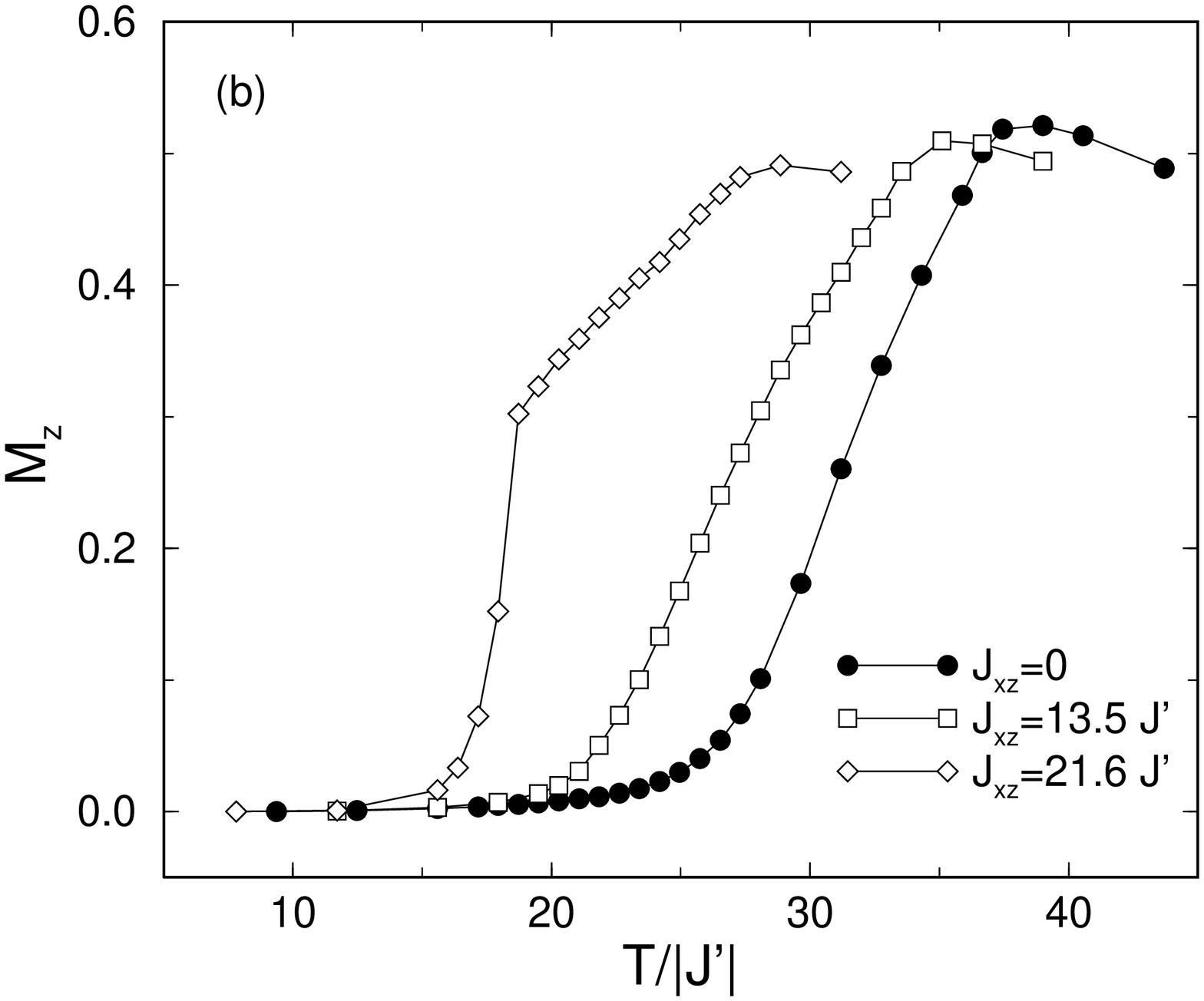,width=7.0cm,angle=270}}
\centerline{\psfig{figure=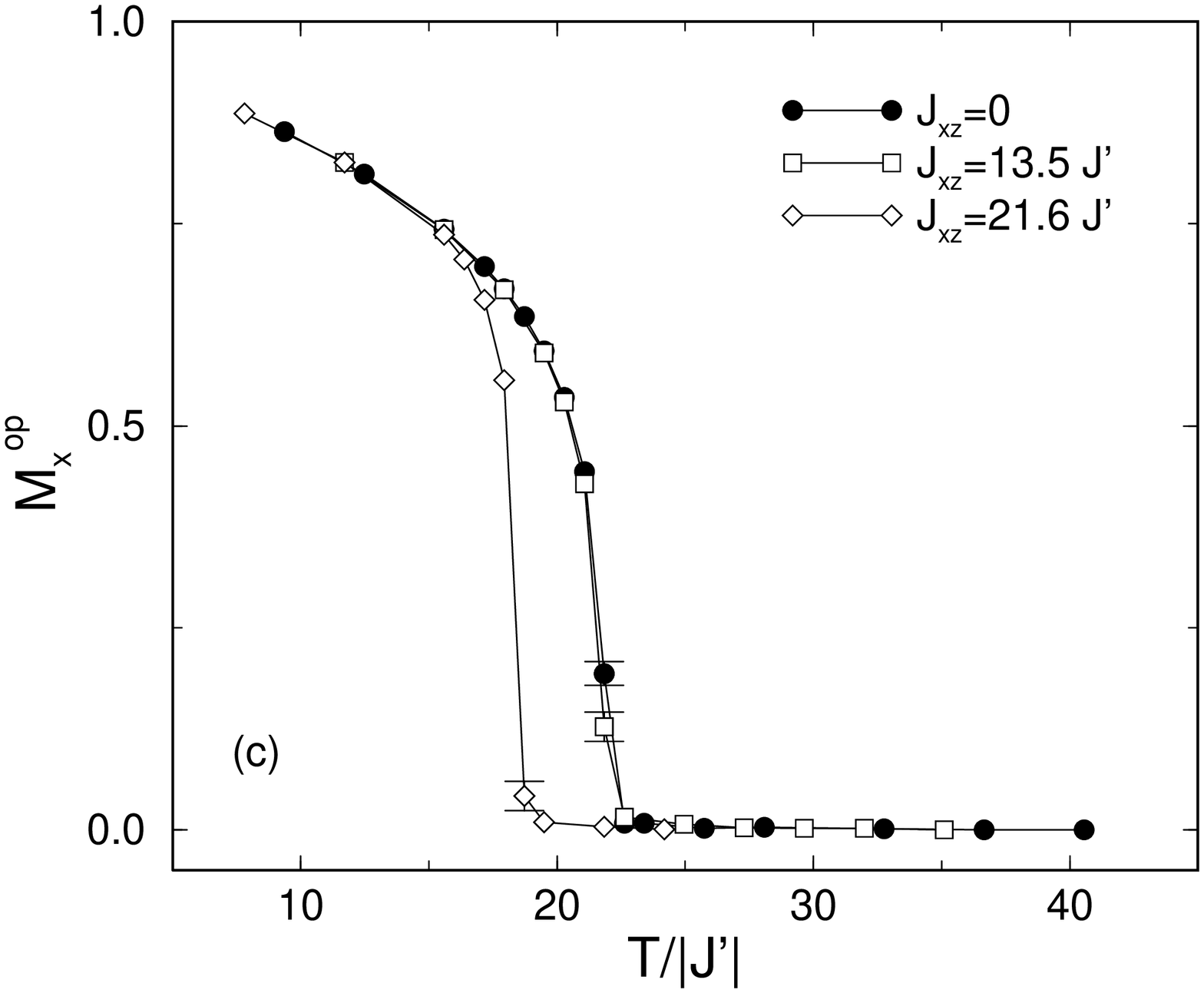,width=7.0cm,angle=270}}
\caption{Temperature dependence of (a) the specific heat (the arrows indicate the
critical temperatures, $T_c$), (b) the total magnetization
in $z$--direction, and (c) the order parameter of the planar spin components for
different values of the nondiagonal spin exchange between the planar and the axial
spin components for the hexagonal antiferromagnetic
model, with $D = -32.4 J'$ and $H_z= -18 J'$. 
Systems with $20 \times 20 \times 20$ spins are considered.
Only error bars larger than the size of the symbols are shown.
}
\label{fig6} \end{figure}

\begin{figure}
\centerline{\psfig{figure=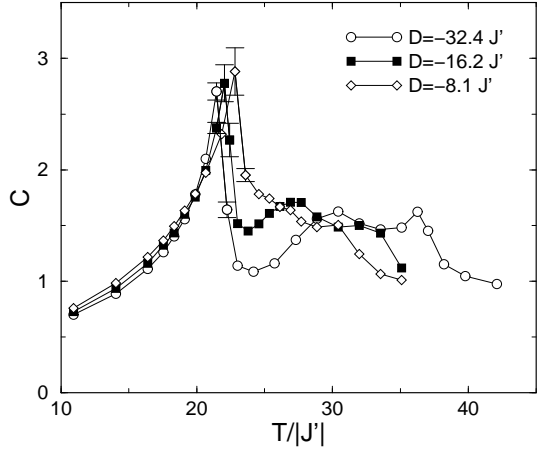,width=7.0cm,angle=270}}
\caption{The specific heat $C$ as a function of temperature for different values
of the single--ion anisotropy, with $J_{xz} = 0$ and $H_z= -18 J'$. The
system size of the simulated antiferromagnetic hexagonal
model is $L =20$. Error bars are only shown when they are larger 
than the sizes of the symbols.
}
\label{fig7} \end{figure}

\begin{figure}
\centerline{\psfig{figure=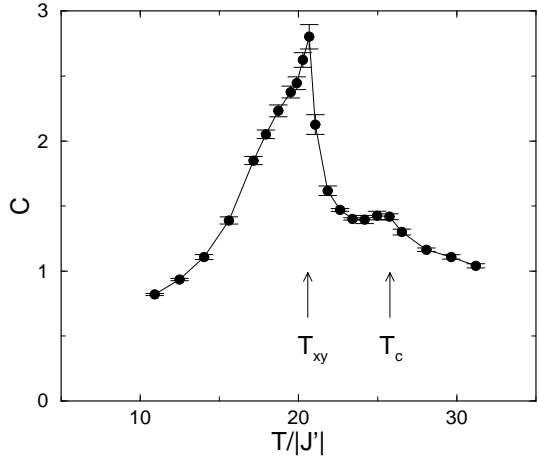,width=7.0cm,angle=270}}
\caption{Temperature dependent specific heat $C$ of the hexagonal
model with $J_{xz} = 16.2 J'$, $D= -8.1 J'$,
and $H_z= -18 J'$, for systems with $20 \times 20 \times 20$ spins.
}
\label{fig8} \end{figure}

\begin{figure}
\centerline{\psfig{figure=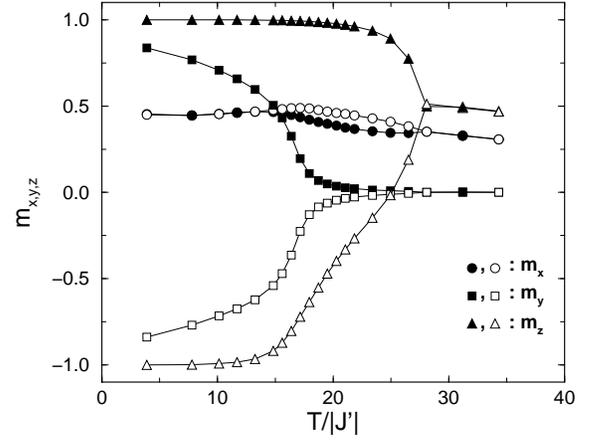,width=7.5cm,angle=270}}
\caption{The components of the
magnetization per layer in odd (full symbols) and even
(open symbols) planes as function of temperature in presence of a planar
field component, $H_x = 0.75 H_z$, with $J_{xz} = 16.2 J'$, $D= -8.1 J'$,
and $H_z= -18 J'$. Hexagonal systems with $30 \times 30 \times 30$ spins were simulated.
}
\label{fig9}
\end{figure}
 
\begin{figure}
\centerline{\psfig{figure=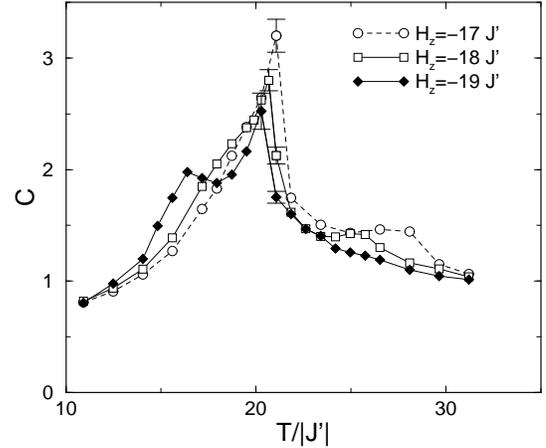,width=7.0cm,angle=270}}
\caption{Specific heat $C$ vs temperature as obtained from Monte Carlo simulations
for different strengths of the axial field, with $J_{xz} = 16.2 J'$, $D= -8.1 J'$,
and $H_x =0$, for hexagonal systems with $20^3$ spins. 
Error bars are only shown when they are larger
than the sizes of the symbols.
}
\label{fig10}
\end{figure}

\end{document}